\documentclass[aps,prl,twocolumn,floatfix,superscriptaddress]{revtex4-1}

\usepackage{amsmath}
\usepackage{amssymb}
\usepackage{times}
\usepackage{braket}
\usepackage[pdftex]{graphicx}
\usepackage{color}

\renewcommand{\vec}[1]{\boldsymbol{#1}}
\newcommand{\change}[1]{\textcolor{black}{#1}}
\newcommand{\changeb}[1]{\textcolor{black}{#1}}
\newcommand{\changec}[1]{\textcolor{black}{#1}}

\makeatletter
\renewcommand\subparagraph{%
 \@startsection {subparagraph}{5}{\z@ }{3.25ex \@plus 1ex
 \@minus .2ex}{-1em}{\normalfont \normalsize \bfseries }}%
\makeatother

\begin{document}
%\captionsetup[figure]{labelfont={bf},name={Fig.},labelsep=period}

\title{Overcoming the speed limit in skyrmion racetrack devices by suppressing the skyrmion Hall effect}

\author{B{\"o}rge G{\"o}bel}
\email[]{bgoebel@mpi-halle.mpg.de}
\affiliation{Max-Planck-Institut f\"ur Mikrostrukturphysik, D-06120 Halle (Saale), Germany}

\author{Alexander Mook}
\affiliation{Institut f\"ur Physik, Martin-Luther-Universit\"at Halle-Wittenberg, D-06099 Halle (Saale), Germany}

\author{J\"urgen Henk}
\affiliation{Institut f\"ur Physik, Martin-Luther-Universit\"at Halle-Wittenberg, D-06099 Halle (Saale), Germany}

\author{Ingrid Mertig}
\affiliation{Institut f\"ur Physik, Martin-Luther-Universit\"at Halle-Wittenberg, D-06099 Halle (Saale), Germany}
\affiliation{Max-Planck-Institut f\"ur Mikrostrukturphysik, D-06120 Halle (Saale), Germany}

\date{\today}

\begin{abstract}
Magnetic skyrmions are envisioned as carriers of information in racetrack storage devices. Unfavorably, the skyrmion Hall effect hinders the fast propagation of skyrmions along an applied electric current and limits the device's maximum operation speed. In this Rapid Communication, we show that the maximum skyrmion velocity increases by a factor of 10 when the skyrmion Hall effect is suppressed, since the straight-line motion of the skyrmion allows for the application of larger driving currents. We consider a ferromagnet on a heavy metal layer, which converts the applied charge current into a spin current by the spin Hall effect. The spin current drives the skyrmions in the ferromagnet via spin-orbit torque. We show by analytical considerations and simulations that the deflection angle decreases, when the spin current is polarized partially along the applied current direction and derive the condition for complete suppression of the skyrmion Hall effect.
\end{abstract}

%\pacs{aaa}

\maketitle

\paragraph{Introduction.} Over the last years, the capacity of data storage devices has steadily grown by reducing the size of magnetic bits in two-dimensional arrays~\cite{parkin2008magnetic}. The minimal size of a bit is limited by quantum effects, so new storage devices have been proposed and tested. Parkin \textit{et al.} suggested a device, that consists of quasi one-dimensional racetracks which can be arranged to create a truly three-dimensional device with a drastically increased storage density~\cite{parkin2004shiftable,parkin2008magnetic,parkin2015memory}. Initially, walls between two ferromagnetic domains were considered as information carriers that can be written, moved, read, and deleted. 

With the discovery of magnetic skyrmions~\cite{skyrme1962unified,bogdanov1989thermodynamically,bogdanov1994thermodynamically, rossler2006spontaneous,muhlbauer2009skyrmion,nagaosa2013topological} the concept of racetrack storage devices could be further improved. In a proposal by Fert \textit{et al.}~\cite{fert2013skyrmions,sampaio2013nucleation} domain walls were replaced by these whirl-like magnetic skyrmions in a ferromagnetic surrounding. Significant advantages of magnetic skyrmions as information carriers are their small size and their topological protection, quantified by an integer topological charge $N_\mathrm{Sk}=\pm 1$.

Besides great stability the nontrivial real-space topology of a skyrmion induces emergent electrodynamic effects: spin-polarized electron currents, injected along the ferromagnetic racetrack, experience a topological Hall effect (THE)~\cite{neubauer2009topological,schulz2012emergent,kanazawa2011large, lee2009unusual,bruno2004topological,hamamoto2015quantized,gobel2017THEskyrmion, gobel2017QHE,gobel2018family} and drive the magnetic skyrmion itself. Detrimentally, this skyrmion propagation is not parallel to the electric current direction; in most scenarios the skyrmion Hall effect (SkHE)~\cite{nagaosa2013topological,zang2011dynamics, iwasaki2013current,jiang2017direct,litzius2017skyrmion, tomasello2014strategy,gobel2018magnetic} limits the maximum velocity, beyond which skyrmions annihilate at the edges of the racetrack. 

To overcome this limitation, several concepts have been established. The combination of two skyrmions with opposite topological charges for example results in antiferromagnetic skyrmions~\cite{zhang2016antiferromagnetic,gobel2017afmskx}, bilayer skyrmions~\cite{barker2016static,zhang2016magnetic} or $2\pi$-skyrmions~\cite{finazzi2013laser,zheng2017direct,zhang2018real,bogdanov1999stability,beg2015ground,beg2017dynamics, zhang2016control,PhysRevB.95.054421}. Their zero topological charge gives a zero SkHE but these quasiparticles have either not been observed experimentally yet or are unstable under motion~\cite{zhang2016control}. 

Another idea is to modify the racetrack setup. Interfacing the actual racetrack with a second ferromagnet, a spin-polarized current can be injected perpendicularly to the interface~\cite{sampaio2013nucleation}. The magnetization direction can be chosen such that the skyrmion moves along the racetrack~\cite{zhang2015skyrmion}. However, the spin current has to be applied over the whole racetrack, what obliterates the necessary low driving currents and the stackability of the racetrack device.

\begin{figure*}
  \centering
  \includegraphics[width=0.95\textwidth]{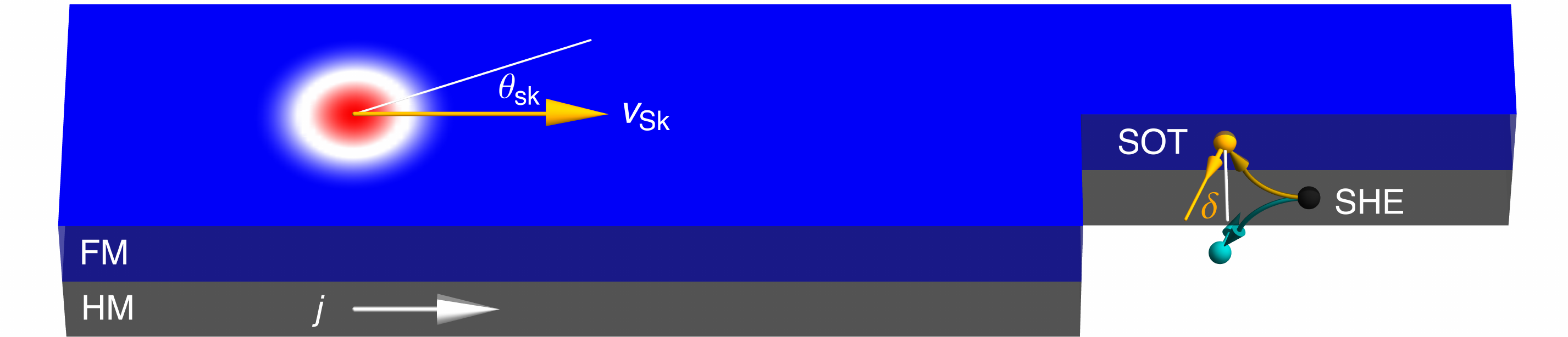}
  \caption{Proposed mechanism to suppress the skyrmion Hall effect. A ferromagnetic layer (FM) is attached to a heavy metal layer (HM), the interfacial DMI stabilizes N\'{e}el skyrmions (red-white circular object) in the FM. When a charge current (density $\vec{j}$) is applied in the HM the spin Hall effect (SHE) generates spins that are injected into the FM where they cause a SOT onto the magnetic moments. In highly symmetric materials the injected spins point into $\pm y$ direction (white line) and the skyrmion moves at a certain skyrmion Hall angle angle $\theta_\mathrm{Sk}$ (white line). In HMs with a reduced symmetry the spin polarization has an angle $\delta$ in the $xy$-plane and the propagation direction of the skyrmion is altered by $\delta$ (orange). The skyrmion Hall effect is absent for $\delta=\theta_\mathrm{Sk}$, see text.}
  \label{fig:racetrack}
\end{figure*}

Replacing the second layer by a nonmagnetic heavy metal (HM) layer, the spin Hall effect (SHE) converts a charge current along the HM into a spin current injected perpendicularly to the ferromagnetic layer (FM)~\cite{sampaio2013nucleation}; see Fig.~\ref{fig:racetrack}. 
In this \changec{Rapid Communication}, we show that the skyrmion Hall angle can be engineered to zero in this setup. For HMs with reduced symmetry the generated spin current is polarized partially along the applied charge current, as recently shown in Refs.~\onlinecite{seemann2015symmetry,wimmer2015spin,zhang2016giant,zhang2017strong}.  This reduces the skyrmion Hall angle compared to a cubic HM layer, where applied charge current, generated spin current and the spin current's polarization are perpendicular to each other. FM and HM materials can be chosen accordingly to suppress the hindering skyrmion Hall effect completely and allow for 10 times as fast skyrmion motion compared to cubic HMs.

\paragraph{Suppression of skyrmion Hall angle.}
The motion of a magnetic skyrmion in a ferromagnetic thin film can be induced via two mechanisms: spin-transfer torque (STT) or spin-orbit torque (SOT). The first mechanism features in-plane injection of a spin-polarized current into the magnetic film. The electron spins align partially with the texture and transfer a torque to the magnetic texture, wherever the latter is non-collinear, in particular at a skyrmion. However, the direction of the injected spins is determined by the texture itself. Therefore the STT scenario does not allow for manipulation of the skyrmion Hall angle (angle of motion with respect to the applied current) and will inevitably suffer from the hindering transverse deflection.

We consider the second mechanism: skyrmion motion via SOT. 
In this scenario the FM is interfaced with a HM layer with a non-vanishing SHE (Fig.~\ref{fig:racetrack}). To propel the skyrmion a charge current $\vec{j}$ is injected along the HM layer ($x$) where the SHE  leads to perpendicular injection ($z$) of spins with polarization
\begin{align}
\vec{s}\propto \sum_{m} \sigma_{zx}^{m} \,\vec{e}_m,
\end{align}
\change{giving a spin Hall angle $\Theta_{\mathrm{SH}}=j^{\vec{s}}_z/j_x$.} $\sigma_{zx}^m$ is an element of the spin conductivity tensor ($x$ applied current direction, $z$ generated spin current direction, $m=\{x,y,z\}$ direction of spin polarization), $\vec{e}_m$ is the unit vector in $m$ direction. For a cubic HM $\sigma_{zx}^{y}\ne 0$ and $\sigma_{zx}^{x}=\sigma_{zx}^{z}=0$, which means that injected spins always point into $\pm y$ direction (we use $-y$ in the following). For this reason skyrmions driven by SOTs via highly-symmetric HMs inhibit to manipulate the skyrmion Hall angle.

However, if the symmetry of the HM allows for $\sigma_{zx}^x\ne 0$, the injected spin orientation becomes $\vec{s}\propto (\sigma_{zx}^{y}\vec{e}_y+\sigma_{zx}^{x}\vec{e}_x)$, with a deviation angle about the $-y$ direction of 
\begin{align}
\delta=\arctan(\sigma_{zx}^{x}/\sigma_{zx}^{y}).
\end{align} 
As we will show, this angle can compensate the hindering skyrmion Hall effect completely.

The collective behavior of the magnetic texture under torques follows the Landau-Lifshitz-Gilbert equation (LLG)~\cite{landau1935theory,gilbert1955lagrangian,slonczewski1996current} 
\begin{align}
\dot{\vec{m}}_i=&-\gamma_e\vec{m}_i\times\vec{B}_{i,\mathrm{eff}}+\alpha\vec{m}_i\times\dot{\vec{m}_i}\label{eq:llg}\\
&+\gamma_e \epsilon\beta[(\vec{m}_i\times\vec{s})\times\vec{m}_i]-\gamma_e \epsilon'\beta(\vec{m}_i\times\vec{s}).\notag
\end{align}
Each magnetic moment $\vec{m}_i$ precesses about the effective magnetic field \changec{$\vec{B}_{i,\mathrm{eff}}=-\delta_{\vec{m}_i}F/M_s$}, which is derived from the \changec{free energy density $F$} (covering exchange, easy-axis anisotropy, interfacial DMI and demagnetization fields). Damping is quantified by the material-dependent Gilbert damping parameter $\alpha$, and $\gamma_e=\gamma/\mu_0=1.760\times 10^{11}\,\mathrm{T}^{-1}\mathrm{s}^{-1}$ is the gyromagnetic ratio of an electron. The in-plane torque coefficient is $\epsilon\beta=\frac{\hbar j\Theta_\mathrm{SH}}{2ed_zM_s}$; \change{we set the out-of-plane torque parameter $\epsilon'$ to zero~\changeb{\footnote{Note, that 'in-plane' and 'out-of-plane' refer to the plane spanned by $\vec{m}_i$ and $\vec{s}$. The 'out-of-plane' torque acts like the torque from a homogeneous magnetic field which is why it does not drive a skyrmion.}}, due to its negligible influence on the skyrmion dynamics~\cite{jiang2017skyrmions,zhang2015skyrmion,
tomasello2014strategy}}.

The generalized Thiele equation~\cite{thiele1973steady} for the SOT scenario describes effectively the center-coordinate motion of a skyrmion~\cite{sampaio2013nucleation}
\begin{align}
b\,\vec{G}\times\vec{v}-b\underline{D}\alpha\vec{v}-B\underline{I}\underline{R}\left(-\delta-\pi/2\right)\vec{j}=0.\label{eq:thielesot}
\end{align}
The skyrmion moves with velocity $\vec{v}$ when driven by the current density $\vec{j}$. The gyromagnetic coupling vector $\vec{G}=G\vec{e}_z$ with \changec{$G=-\int\vec{m}(\vec{r})\cdot[\partial_{x}\vec{m}(\vec{r})\times\partial_{y}\vec{m}(\vec{r})]\,\mathrm{d}^2r=-4\pi N_\mathrm{Sk}$} and the dissipation tensor $\underline{D}$ with \change{$D_{ij}=\int\partial_{x_i}\vec{m}(\vec{r})\cdot\partial_{x_j}\vec{m}(\vec{r})\,\mathrm{d}^2r$} (only $D_{xx}$, $D_{yy}$ non-zero) determine the motion direction.
$\underline{R}(\phi)$ is a rotation matrix in the $xy$ plane around the angle $\phi$, $\delta$ characterizes the injected spin orientation with respect to the $-y$ direction, and $B=\hbar/(2e)\Theta_\mathrm{SH}$~\cite{sampaio2013nucleation,hrabec2017current}. The factor $b=M_s d_z/\gamma_e$ contains the thickness of the FM $d_z$ and its saturation magnetization $M_s$. $\underline{I}$ is the tensor \change{$I_{ij}=\int[\partial_{x_i}\vec{m}(\vec{r})\times\vec{m}(\vec{r})]_{x_j}\,\mathrm{d}^2r$}. A N\'{e}el skyrmion (helicity $h=0$) has only $I_{xy}$ and $I_{yx}$ nonzero. For a different helicity this particular tensor is rotated by $-h$.

\begin{figure*}
  \centering
  \includegraphics[width=0.9\textwidth]{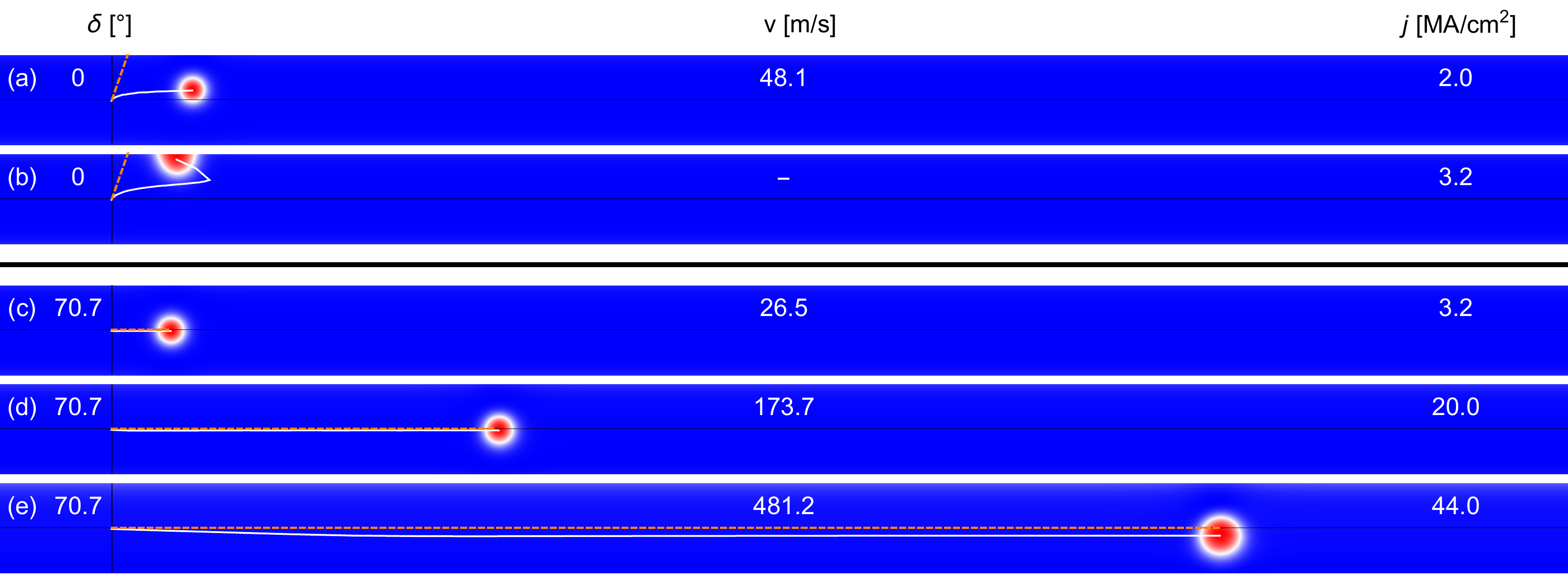}
  \caption{Micromagnetic simulations comparing conventional and optimized setup. (a,b) Conventional CoPt racetrack with $\delta=0^\circ$. (c-e) Optimized racetrack with $\delta=70.7^\circ$. Results of the simulations are shown after 1 ns propagation duration at different current densities $\Theta_\mathrm{SH}j$ (indicated on the right). The optimal geometry yields a stable motion for much larger current densities, allowing for faster propagation of the skyrmion. The skyrmion's velocity in (e) is 10 times as large as in the conventional scenario in (a); $v=481.2\,\mathrm{m}/\mathrm{s}$ compared to $v=48.1\,\mathrm{m}/\mathrm{s}$. Red: $-z$, blue: $+z$ magnetization; white shows the actual trajectory and dashed orange shows the predicted trajectory from the Thiele equation without confining potential. An animated version of this figure is presented as Supplementary material.}
  \label{fig:delta0}
\end{figure*}

The Thiele equation gives a skyrmion Hall angle \change{$\theta_\mathrm{Sk}=\arctan(v_y/v_x)$} which is zero for 
\begin{align}
\tan(\delta+h)=\frac{G}{\alpha D_{xx}},
\end{align}
determined via $v_y=0$.
This tells that an optimal spin orientation $\delta$ or an optimal skyrmion helicity $h$ can be found, for which the skyrmion Hall angle is absent. For $\delta=0$, the above equation condenses the recent geometric considerations by Kim \textit{et al.}~\cite{kim2018asymmetric} to a simple expression. They showed that for a mixed interfacial and bulk Dzyaloshinskii-Moriya interaction~\cite{dzyaloshinsky1958thermodynamic,moriya1960anisotropic} (DMI), i.\,e., $h\ne 0,\pi/2$, the skyrmion Hall angle can vanish. However, such skyrmions are still awaiting experimental identification.

Due to interfacial DMI the SOT scenario typically stabilizes N\'{e}el skyrmions ($h=0$). A vanishing skyrmion Hall angle is then achieved when the skyrmion Hall angle $\arctan[G/(\alpha D_{xx})]$ of the system with $\sigma_{zx}^x=0$ is compensated by $\delta=\arctan(\sigma_{zx}^x/\sigma_{zx}^y)$:
\begin{align}
\frac{\sigma_{zx}^x}{\sigma_{zx}^y}=\frac{G}{\alpha D_{xx}}.\label{eq:delta0}
\end{align}
\changeb{The effective torque has been manipulated via the injected spin orientation instead of the magnetic texture itself, so that the skyrmion Hall angle is completely suppressed.}

\paragraph{High-speed skyrmions in micromagnetic simulations.}

\begin{figure}[h!]
  \centering
  \includegraphics[width=\columnwidth]{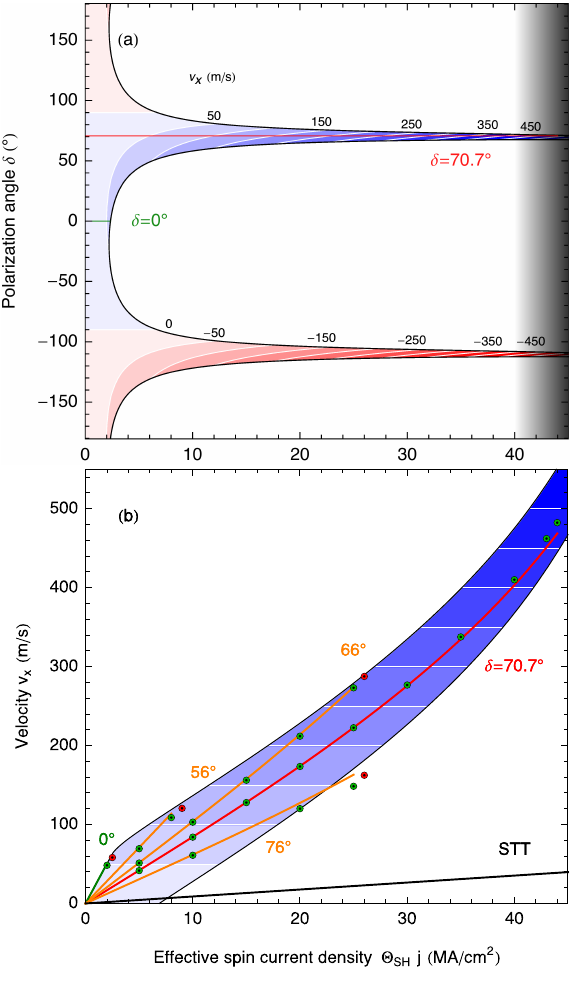}
  \caption{Skyrmion velocity and stability for different HM materials. (a) The range of allowed $\delta$ directions over $j$ (colored area). (b) The $j$ dependence of the velocity for selected $\delta$.  
The color in both panels corresponds to the velocity along the racetrack (blue positive, red negative; indicated with numbers). The colored lines correspond to constant $\delta$ values (indicated). Lines and velocities have been calculated from the Thiele equation, \change{without considering pinning effects which become negligible at large current densities}.
Explicit results from the micromagnetic simulations are indicated by points (green: stable motion, red points: skyrmion annihilation). The black line (STT) is taken from Ref.~\onlinecite{sampaio2013nucleation} for comparison.
}
\label{fig:vel}
\end{figure}

We verify the results of the Thiele equation using micromagnetic simulations and begin with CoPt, as considered in Ref.~\onlinecite{sampaio2013nucleation}, where $-y$ polarized spins are injected into the FM.
We use Mumax3~\cite{vansteenkiste2011mumax,vansteenkiste2014design} to solve the LLG equation with the SOT term~\eqref{eq:llg}. The racetrack geometry throughout this \changec{Rapid Communication} is: width $40\,\mathrm{nm}$, \changec{thickness $d_z=0.5\,\mathrm{nm}$}, periodic boundary conditions in $x$ direction, and a cell size of $0.5\,\mathrm{nm}\times 0.5\,\mathrm{nm}$.
The parameters for CoPt read~\cite{sampaio2013nucleation}: exchange stiffness $J=15\,\mathrm{pJ}/\mathrm{m}$, DMI constant $D=3\,\mathrm{mJ}/\mathrm{m}^2$, uniaxial anisotropy $K_z=0.8\,\mathrm{MJ}/\mathrm{m}^3$, saturation magnetization $M_s=0.58\,\mathrm{MA}/\mathrm{m}$ and Gilbert damping $\alpha=0.3$. Note, that the choice of parameters does not \emph{qualitatively} affect the skyrmion motion \changec{(as long as individual skyrmions can be stabilized). This is especially important since ab-initio calculations predict $D$ values that differ by up to 60\%~\cite{simon2018magnetism}}.

When driven by SOT a stabilized N\'{e}el skyrmion is pushed in $x$ direction (along the track) via the dissipative force (proportional to $D_{xx}$) and is also pushed in $y$ direction (to the edge) by the gyrotropic force (proportional to $N_\mathrm{Sk}$).

The confined geometry introduces another term to the right side of the Thiele equation~\eqref{eq:thielesot}, which is the gradient $\nabla U(y)$ of the potential energy of the skyrmion in the ferromagnetic layer. It contributes strongly at the edge of the racetrack, repels and slightly deforms a skyrmion. Using the parameters of CoPt N\'{e}el skyrmions are stabilized with $D_{xx}=14.63$ and $I_{xy}=55.58\,\mathrm{nm}$ what gives a skyrmion Hall angle of about $70.7^\circ$. The skyrmions are not perfectly rotational symmetric. 

In Fig.~\ref{fig:delta0}a the skyrmion is driven by a considerably low current density of $\Theta_\mathrm{SH}j=2\,\mathrm{MA}/\mathrm{cm}^2$. In the beginning the skyrmion moves under the above angle. Increasing the $y$ coordinate upon propagation leads to an increase of the repelling force from the confining potential. After this short acceleration phase the skyrmion moves in a steady state along the track. For current densities beyond a threshold, the gyrotropic force in $y$ direction is so strong, so that the confining potential is overcome and the skyrmion annihilates at the edge of the racetrack (Fig.~\ref{fig:delta0}b). In this particular CoPt racetrack \changeb{the applicable current density is limited to about $\Theta_\mathrm{SH}j= 3\,\mathrm{MA}/\mathrm{cm}^2$ what limits} the velocity to about 50\,m/s.

Next, we consider a HM material with nonzero $\sigma_{zx}^x$, i.\,e., when an electric current is applied along the racetrack ($x$) the injected spins are not polarized in $-y$ direction but are rotated by $\delta$. Choosing the tensor elements so that equation~\eqref{eq:delta0} is fulfilled, the skyrmion Hall effect is completely suppressed and the skyrmion moves in the middle of the racetrack (Fig.~\ref{fig:delta0}c). The trajectory prediction from the Thiele equation coincides with the simulated trajectory.

The velocity of this optimized racetrack is less than that of the CoPt racetrack for the same $j$ which follows directly from the Thiele equation, 
\begin{align}
v_x=\frac{B}{b}\frac{I_{xy}\cos\delta}{\alpha D_{xx}}j.\label{eq:vmax}
\end{align}
However, skyrmions in this optimized geometry can be driven by currents 20 times as large and reach skyrmion velocities 10 times as large as for the CoPt racetrack. The result of a micromagnetic simulation for a fast-moving skyrmion is depicted in Fig.~\ref{fig:delta0}e. 

We showed that the skyrmion Hall angle vanishes in the optimized setup. However, $\delta$ is given by material-specific parameters that can be tuned only slightly. Therefore, we analyze the range of current densities yielding a stable motion in dependence of $\delta$ and find that the maximum skyrmion velocity (at maximal allowed current density) increases in every case compared to the $\sigma_{zx}^x=0$ velocity, even if $\delta$ deviates from the optimal value.

The skyrmion motion is stable in the colored areas of Fig.~\ref{fig:vel}a (solved analytically in Supplementary material 1). For very low current densities the skyrmion remains stable for every spin polarization orientation since the transverse force is too weak to overcome the confining potential regardless of the propagation direction. For large $j$ the skyrmion can only `survive' if the skyrmion Hall angle is zero. The motion is reversed for $\delta\rightarrow\delta+180^\circ$ (sign reversal of current direction or spin conductivity tensor elements).

The velocity along the racetrack (represented by the color scale in Fig.~\ref{fig:vel}) is in first approximation proportional to the applied current density $j$ [cf. equation~\eqref{eq:vmax}]. Yet $I_{xy}$ and $D_{xx}$ depend on $j$ as well. A power-law fit of these quantities (see Supplementary material 2) correctly reproduces the trend of the micromagnetic results (Fig.~\ref{fig:vel}b). Not only does the skyrmion velocity $v_x$ increase with current density but also the efficiency $v_x/j$. 

The $j$ dependence of $D_{xx}$ also affects the optimal $\delta$ value [cf. equation~\eqref{eq:delta0}]. It leads to the emergence of a small skyrmion Hall effect for $\delta=70.7^\circ$ at high current densities (clearly visible in the simulation in Fig.~\ref{fig:delta0}e), while being negligible for lower current densities (Fig.~\ref{fig:delta0}d). The optimal $\delta$ shifts to about $69^\circ$ for $\Theta_\mathrm{SH}j= 44\,\mathrm{MA}/\mathrm{cm}^2$. In any case the maximum applicable current density is limited to $\Theta_\mathrm{SH}j\approx 45\,\mathrm{MA}/\mathrm{cm}^2$. Even for the corrected $\delta$ the skyrmion becomes so extended that it touches both edges of the racetrack. This limit of the maximal velocity therefore depends on the racetrack's width (here $40\,\mathrm{nm}$).

The results of the micromagnetic simulations confirm the predictions of the Thiele equation quantitatively well, especially for the optimal $\delta$. Small deviations of analytical and numerical results are attributed to the fact that $D_{xx}$ and $I_{xy}$ are not only $j$ but also $\delta$ dependent; a skyrmion at the edge of the racetrack has a different shape compared to a skyrmion in the middle of the racetrack. For Fig.~\ref{fig:vel} we fitted only the $j$ dependence for a fixed $\delta=70.7^\circ$ which already yields good agreement with the simulations.

\paragraph{Discussion.}
In the previous Section, we proved that the skyrmion Hall angle can be suppressed by a nonzero $\sigma_{zx}^x$. To utilize the demonstrated advantage of SOT-driven skyrmions on a racetrack a HM with a nontrivial spin conductivity tensor has to be used. It has to generate a spin current with spins partially oriented along the charge current direction.

This nonzero $\sigma_{zx}^x$ can be realized in nonmagnetic materials for triclinic, monoclinic, trigonal and in tetragonal/hexagonal crystal systems, if the latter two have C$_4$, S$_4$ and C$_{4\mathrm{h}}$ symmetry. Besides prohibition in all other tetragonal/hexagonal crystal systems, the element is forbidden to arise in orthorombic and cubic crystal systems (cf. symmetry analysis of the spin conductivity tensor in Ref.~\onlinecite{seemann2015symmetry}). 

The results of Ref.~\onlinecite{wimmer2015spin} suggest that Pt$_3$Ge, Au$_4$Sc and (Au$_{1-x}$Pt$_x$)Sc may be suitable candidates for our predicted HM layer setup: they have a non-zero $\sigma_{zx}^x$ element; e.\,g., for (Au$_{0.8}$Pt$_{0.2}$)Sc the authors of that publication find $\sigma_{zx}^x/\sigma_{zx}^y=1/3$. Moreover, these materials are non-magnetic and consist of elements with a sizable spin-orbit coupling, which is expected to constitute interfacial DMI necessary for stabilizing N\'{e}el skyrmions. 
Even though unfavorable as a HM layer, magnetic Mn$_3$Rh, Mn$_3$Ir and Mn$_3$Pt \change{were} shown to have $\sigma_{zx}^x/\sigma_{zx}^y$ of $1.903$, $1.288$ and $2.063$, respectively~\cite{zhang2017strong,zhang2016giant}. 

There exist many materials in the above crystal systems, which can potentially have even larger ratios. Even if a setup does not fulfill the optimal condition, the merest rotation of the injected spin orientation improves the racetrack.

Summarizing, we have presented an approach to suppress the skyrmion Hall effect in a bilayer system via spin-orbit torques. Relation~\eqref{eq:delta0} comprises the parameters that determine the magnitude of the skyrmion Hall angle in a spin-orbit torque scenario: Gilbert damping $\alpha$, skyrmion shape $G/D_{xx}$, and injected spin polarization angle $\delta$. While the magnetic layer provides the Gilbert damping $\alpha$, the heavy-metal layer determines the quotient of spin Hall conductivity tensor elements $\delta$. The two materials have to be chosen accordingly to ensure a suppression of the SkHE.

At the current state of skyrmion racetrack memory development, our proposal is likely more feasible to tune the skyrmion Hall angle to zero compared to the concepts presented in the introduction. 
%It features very stable N\'{e}el skyrmions, an in-plane injection of low current densities and an increased propagation velocity along the racetrack (factor of 10 compared to conventional setups). 
We demonstrated that by suppressing the skyrmion Hall effect skyrmion velocities of up to 500\,m/s can be achieved at a reasonable $v_x/j$ efficiency. 
%The velocity can be further increased by selected heavy metals (with smaller optimal $\delta$ values) and selected racetrack geometries (altering the energy potential and the critical edge-repulsion force).

\begin{acknowledgments}
  This work is supported by Priority Program SPP 1666 and
SFB 762 of Deutsche Forschungsgemeinschaft (DFG).
\end{acknowledgments}

\bibliography{short,MyLibrary}
\bibliographystyle{apsrev}

\end{document}